# The criterion of planar instability in alloy solidification under varying conditions: A viewpoint from free energy

Fengyi Yu

CAS Key Laboratory of Mechanical Behavior and Design of Materials, Department of Modern Mechanics, University of Science and Technology of China, Hefei 230026, China

E-mail address: fengyi.yu.90@gmail.com

**ABSTRACT**

In alloy solidification, the transport processes of heat and solute result in morphological instability of the interface, forming different patterns of solidification structure and determining the mechanical properties of components. As the first observable phenomenon of the morphological instabilities, the planar instability influences the subsequent stages significantly, deserving in-depth investigations. In this paper, the planar instability in alloy solidification under varying conditions is studied. Firstly, the dynamical evolution of the planar instability is performed by the theoretical model and phase-field model, respectively. Secondly, to represent the history-dependence of solidification, the varying parameters are adopted in the simulations. Then the criterion of the planar instability under the varying conditions is discussed. This paper considers the critical parameters of the planar instability are the excess free energy at the interface and corresponding interfacial energy. Finally, to validate the criterion, the comparisons between the phase-field model and theoretical model are carried out, showing good consistency. Moreover, solidification processes with different preferred crystallographic orientations are performed, demonstrating the effect mechanism of the excess free energy and interfacial energy on the planar instability. The idea of the interfacial energy influencing the planar instability could be applied to investigating other patterns induced by interfacial instability.

# I. INTRODUCTION

The solidification structures dominate the mechanical properties of as-solidified parts. An accurate prediction of the solidification structures could provide a theoretical basis for optimizing the parameters. To achieve the accurate prediction, the solidification dynamics need to be revealed. Due to the different characteristics of physical processes at different scales, the investigation of solidification dynamics has been a long-standing challenge. [1,2] From the viewpoint of mesoscale, solidification structures are dominated by the interactions between interfacial processes and transport processes of heat and solute. [3-5] The diffusive nature of the transport processes, including spatial and temporal evolution, gives rise to morphological instability of the solid/liquid (S/L) interfaces, resulting in different patterns of solidification structure.

As the first observable phenomenon in the evolution of solidification patterns, the planar instability affects the subsequent solidification stages significantly [6,7], deserving in-depth investigations. Chalmers et al. [8] analyzed the heat and solution balance at a moving S/L interface. They gave the idea of Constitutional Supercooling (CS) and established the CS criterion of the morphological instability by $G/V_P < \Delta T_0 / D_L$. In the expression, $G$ is the thermal gradient, $V_P$ is the pulling speed, $\Delta T_0$ is the solidification temperature range in the phase diagram, and $D_L$ is the solute diffusion coefficient in the liquid. Although it reveals the thermodynamic essence of the interfacial instability, the CS theory does not account for the transport processes of heat and solute. Mullins and Sekerka [9,10] analyzed the stability of a crystal, based on a dynamic approach in which the equations governing heat flow and solute diffusion are solved simultaneously while allowing for a change of shape due to a perturbation, known as the MS theory. Compared with the CS theory, the MS theory is based on a dynamic approach, considering the interplay of the diffusion transport and interfacial energy, representing the spatial and temporal evolution of solidification patterns. However, the MS theory is performed assuming a steady-state planar interface, neglecting the time dependence of diffusion transport. By assuming the solute concentration evolves with time, Warren and Langer (WL) [11] extended the MS theory to non-steady-state dynamics. Their analysis indicates the solidification evolution is history-dependent, depending on the detailed way in which the sample is prepared and set in motion. The interfacial instability predicted by the WL model agrees well with the experimental observations of real-time synchrotron X-ray radiography [12], demonstrating the validity of the WL model. Recently, by combining the time-dependent linear stability analysis in the WL model with the Fourier synthesis, Wang et al. [13] developed a simple model to predict the morphological evolution of the S/L interface directly at the initial stage. The model is verified by the experimental observations under the steady-state conditions [14]. Subsequently, Dong et al. [15,16] modified this model from steady-state conditions to non-steady-state conditions, i.e., the model can represent the time-dependent G and $V_P$. The onsets of the planar instability under the same G and $V_P$, with different increasing rates of $V_P$, were carried out. The results illustrate the increasing rate of $V_P$ does affect the solidification evolution, including the incubation time and average wavelength of the planar instability. This study demonstrates the microstructure evolution depends on the detailed way the solidification conditions are achieved. To date, considerable investigations of the interfacial instability have been made. However, these theoretical models involve many approximations and simplifications, resulting from the constraint that the solutions of analytical and semi-analytical models can only apply under simple conditions. As a result, these theoretical models could hardly handle the complex morphologies of the S/L interfaces and the corresponding interfacial effects. Moreover, the as-simplified solidification conditions are far from the realistic processes, limiting the application of these theoretical models.

Compared with analytical methods, numerical methods could solve the equations under complex conditions, having the advantage of representing relatively realistic processes. As a representative, the Phase-Field (PF) method combines the insights of thermodynamics and the dynamics of transport processes, having solid physical foundations. [17-20] Moreover, since it avoids the shape error caused by tracking interfaces during computation, the PF method has high numerical accuracy. [18-21] By introducing a phenomenological "Anti-Trapping Current" (ATC) term [22], the PF model can simulate alloy solidification quantitatively. The quantitative PF model has been applied to increasingly complex conditions, from isothermal solidification [22], directional solidification [23,24] to melt pool solidification. [25,26] Based on the PF simulations, the mechanisms of solidification evolution are studied, including the planar to cellular transition [14,15,27] the selection of growth direction [28-30], the competitive growth [31-33], the columnar to equiaxed transition [34,35], and the sidebranching dynamics [36-38], etc. The PF simulations agree well with the experimental observations, indicating the accuracy of the PF method. In one word, the PF model avoids the limitations caused by the simplifications in the analytical models. Moreover, since it can capture the complex morphologies and characteristic parameters of the interfaces, the PF model can represent the interplay between the interfacial processes and transport processes accurately, which is suitable for investigating the interfacial instability systematically.

In this paper, the planar instability in alloy solidification under varying conditions is studied. Firstly, the evolution of the planar instability is represented by the WL model and PF model, respectively. Secondly, to represent the history-dependence of solidification, the dynamic parameters are adopted in the simulations. Based on the simulations, the criterion of the planar instability under varying conditions is discussed. This paper considers the critical parameters of the instability are the excess free energy at the S/L interface and the corresponding interfacial energy. Finally, to validate the criterion, the comparisons between the PF model and theoretical model are carried out. Moreover, solidification processes with the different Preferred Crystallographic Orientations (PCOs) are performed, demonstrating the influences of the excess free energy and interfacial energy on the planar instability.

## II. MODELS AND METHODOLOGY

In directional solidification, the so-called "frozen temperature approximation" is adopted, given by:

$$T(z,t) = T_0 + G\left(z - z_0 - \int V_P(t)dt\right) \quad (1)$$

where $T_0$ is the melt temperature of the pure material, the pulling direction is along the z axis, and $z_0$ is the position of the interface. G is the temperature gradient, and $V_P$ is the pulling speed. This approximation is on the basis of the assumptions: (1) The latent heat is ignored, i.e., the temperature field is undisturbed by the evolution of the S/L interface. It is essentially a statement concerning the relative magnitudes of the terms in the Stefan condition, $\rho_s L_f v^*_n \ll k_{s,l} \nabla T_{s,l} \cdot \mathbf{n}$. In the expression, $\rho_s$ is the density of the solid phase, $L_f$ is the latent heat, and $v^*_n$ is the rate of solidification. $k_{s,l}$ means the thermal conductivity. $\nabla T_{s,l}$ is the thermal gradient, and $\mathbf{n}$ is the normal direction of the S/L surface. The rate of alloy solidification ($v^*_n$) is limited by solute diffusion. Since the coefficient of thermal conductivity is much larger than solute diffusion, the crystallization latent heat can be released quickly through heat conduction. As a result, the effect of latent heat on the temperature field could be ignored. [39] (2) There is no flow in the liquid, consistent with the assumption that the densities of the solid and liquid are equal. [3] It should be pointed out, the frozen temperature approximation is just for the heat transport. The thermodynamic model for solidification contains the latent heat, both in the theoretical model and the PF model.

### A. Theoretical model

#### 1. Description of the model

The theoretical model is based on the linear instability analysis under the non-steady-state conditions, the detailed derivations can be found in literature [11,13,14]. The following are the key equations.

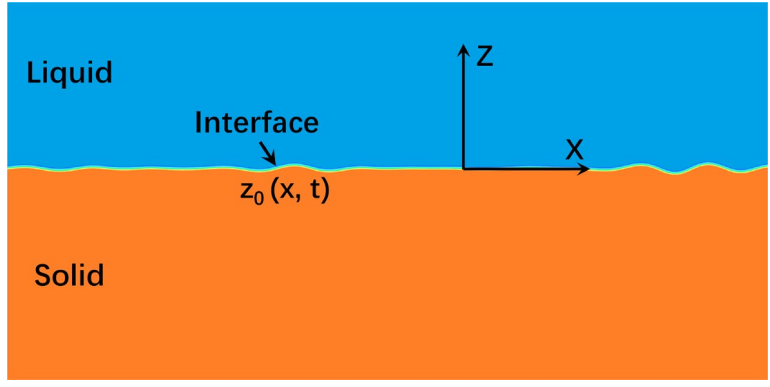

Fig. 1. The sketch of the coordinate system at the initial growth stage

The sketch of the coordinate system at the initial stage of directional solidification is shown in Fig. 1, where the pulling direction of the interface is along the z axis. Assuming local equilibrium at the interface, the concentration ahead of the planar front is:

$$c_0(z_0, t) = -\frac{G}{m} \cdot z_0(t) \quad (2)$$

where m is the slope of the liquidus line in the phase diagram. $z_0$ is the position of the interface, based on which the instantaneous velocity of the interface is defined as: [11]

$$V_I = V_P(t) + \frac{\partial z_0}{\partial t} \quad (3)$$

where $V_P$ is the time-dependent pulling speed. At the interface, solute should satisfy the conservation law:

$$-D_L \left.\frac{\partial c_0}{\partial z}\right|_{z0} = V_I(1-k)c_0(z_0, t) \quad (4)$$

where $D_L$ is the solute diffusion coefficient in the liquid, and k is the solute partition coefficient. Furthermore, at the planar growth stage, the time-dependent concentration can be approximated by the expression: [11]

$$c_0(z, t) = c_\infty + \left[c_0(z_0, t) - c_\infty\right] \cdot \exp\left[\frac{-2(z - z_0)}{l}\right] \quad (5)$$

where $c_\infty$ is the average concentration, and l is the diffusion length. The partial derivative of equation (5) is:

$$\frac{\partial c_0}{\partial z} = \left[c_0(z_0, t) - c_\infty\right] \cdot \exp\left[\frac{-2(z - z_0)}{l}\right] \cdot \left(\frac{-2}{l}\right) \quad (6)$$

Taking equation (6) to equation (4):

$$\left.\frac{\partial c_0}{\partial z}\right|_{z0} = \left[c_0(z_0, t) - c_\infty\right] \cdot \left(\frac{-2}{l}\right) = \frac{V_I(1-k)c_0(z_0, t)}{-D_L} \quad (7)$$

Then, the time-dependent concentration at the interface can be expressed as:

$$c_0(z_0, t) = \frac{2D_L c_\infty}{2D_L - V_I(1-k)l} \tag{8}$$

Finally, using the local equilibrium condition in equation (2), the time-dependent interface position $z_0$ and diffusion length $l$ could be expressed as: [11]

$$\frac{\partial z_0}{\partial t} = V_I - V_P(t) = \frac{2D_L(z_0 - z_\infty)}{l(1-k)z_0} - V_P(t) \tag{9}$$

$$\frac{\partial l}{\partial t} = \frac{4D_L(z_\infty - kz_0)}{l(1-k)z_0} - \frac{l}{z_0 - z_\infty}\frac{\partial z_0}{\partial t} \tag{10}$$

where $z_\infty$ is the steady-state position of the planar interface with the relation of $z_\infty = -m \cdot c_\infty/G$.

Based on the time-dependent linear stability analysis and the assumption of an infinitesimal sinusoidal perturbation with spacing frequency ω, the increase rate of perturbation amplitude can be given by:

$$\sigma_\omega(t) = [dA_\omega(t)/dt]/A_\omega(t) \tag{11}$$

where $A_\omega(t)$ is the amplitude of the fluctuation.

Combining the time-dependent interface position and diffusion length in equations (9)-(10), the linear stability analysis of the accelerating planar interface yields the dispersion relation of the perturbation under the transient conditions, as shown in equation (12):

$$q_\omega \left\{ 1 + \frac{2[z_0(t) - z_\infty]}{l(t)} + \frac{\Gamma\omega^2}{G} \right\} = \\ \frac{V_I(t) - V_P(t)}{D_l} + \frac{2[z_0(t) - z_\infty]}{l(t)} \cdot \left[ \frac{V_I(t)}{D_l} + \frac{\sigma_\omega(t)}{V_I(t)} + \frac{1}{z_0(t)} + \frac{\Gamma\omega^2}{G \cdot z_0(t)} \right] \tag{12}$$

where $q_\omega$ is the inverse decay length of the concentration fluctuation at the interface along the z direction. $\Gamma$ is the Gibbs-Thomson coefficient.

Based on equation (12), the time-dependent increase rate of the amplification rate $\sigma_\omega(t)$ can be obtained. Then, according to the solution of equation (11), the time-dependent amplitude is given by:

$$A_\omega(t) = A_\omega(0) \exp\left[ \int_{t_0}^{t} \sigma_\omega(t) dt \right] \tag{13}$$

where $t_0$ is the critical time when $\sigma_\omega$ changes from negative to positive, and $A_\omega(0)$ is the initial amplitude of the infinitesimal fluctuation.

## 2. Simulation procedure

By solving equations (8)-(10), the important characteristic parameters can be obtained, including the time-dependent position of the S/L interface $z_0$, the concentration ahead of the interface $c_0$, the instantaneous interfacial velocity $V_I$ and the diffusion length ahead of the planar front $l$.

For a small time interval $\Delta t$, the relations in equations (14)-(15) can be regarded as the initial conditions when solving equations (8)-(10) numerically. [11]

$$l \approx \left( \frac{8 D_L \cdot \Delta t}{3} \right)^{1/2} \tag{14}$$

$$z_0 = z_\infty - V_P(t) \cdot \Delta t + \frac{V_P(t)\sqrt{2 D_L}}{\sqrt{3} \cdot z_\infty (1-k)} (\Delta t)^{3/2} \tag{15}$$

When solving equation (12), the dispersion relation, the increase rate $\sigma_\omega$ satisfies equation(16): [11]

$$\sigma_\omega = D_l \left( q_\omega^2 - \omega^2 \right) - q_\omega V_I \tag{16}$$

Setting $\sigma_\omega$=0, the solution of equation (16) is:

$$q_\omega = \frac{V_I}{2 D_l} + \sqrt{\omega^2 + \left( \frac{V_I}{2 D_l} \right)^2} \tag{17}$$

Eliminating $q_\omega$ from these equations and inserting the values of $z_0$ and $l$ from the previous calculation, a time-dependent spectrum of the amplification increase rate $\sigma_\omega(t)$ can be obtained. Taking $\sigma_\omega(t)$ to equation (13), the time-dependent amplitude can be obtained. In equation (13), $A_\omega(0)$ is the initial amplitude. Based on the approximation of equilibrium fluctuation spectrum, $A_\omega(0)$ equals to the capillary length $d_0$. [13]

## B. Phase-Field model

### 1. Description of the model

The detailed derivation and validation of the quantitative PF model for alloy solidification can be found in the literature [22-24]. The following is a brief introduction of this PF model, presenting the equations describing the evolution of phase field and solute field.

Firstly, a scalar variable $\phi(\mathbf{r}, t)$ is introduced to identify the phase, where $\phi$=+1 reflects the solid phase, $\phi$=−1 reflects the liquid phase, and intermediate values of $\phi$ correspond to the S/L interface. Since $\phi$ varies smoothly across the interface, the usual sharp interface becomes diffuse, and the phases turn into a continuous field, i.e., the phase field.

For the solute field, the composition c(**r**, t) is represented via the supersaturation field U(**r**, t):

$$U = \frac{1}{1-k}\left(\frac{2kc/c_\infty}{1+k-(1-k)\cdot\phi} - 1\right) \tag{18}$$

where k is the solute partition coefficient, $c_\infty$ is the average solute concentration.

For alloy solidification, the ATC term could recover the local equilibrium at the S/L interface. Moreover, the ATC term could eliminate the spurious effects when the interface width is larger than the capillary length. The ATC term is given by: [22,24]

$$\vec{j}_{at} = -\frac{1}{2\sqrt{2}}\left[1+(1-k)U\right]\frac{\partial\phi}{\partial t}\frac{\vec{\nabla}\phi}{|\vec{\nabla}\phi|} \tag{19}$$

where ∂ϕ/∂t means the rate of solidification. ∇ϕ/|∇ϕ| is the unit length along the normal direction of the S/L interface.

For the cubic crystal Al-Cu alloy, a four-fold anisotropy function in 2D system is used in this paper:

$$a_s(\hat{n}) \equiv a_s(\theta+\theta_0) = 1 + \varepsilon_4 \cos 4(\theta+\theta_0) \tag{20}$$

where $\varepsilon_4$ is the anisotropy strength, θ the angle between the normal direction of S/L interface and the pulling direction (here is the z-axis), $\theta_0$ is the intersection angle between the PCO of grain and the z-axis.

Finally, the governing equations of phase field and supersaturation field are given by: [22-24]

$$a_s^2(\hat{n})\left[1-(1-k)\frac{z-z_0-\int V_P(t)dt}{l_T}\right]\frac{\partial\phi}{\partial t} =$$
$$\nabla\cdot\left[a_s^2(\hat{n})\vec{\nabla}\phi\right] - \partial_x\left(a_s(\hat{n})\cdot a_s'(\hat{n})\cdot\partial_y\phi\right) + \partial_y\left(a_s(\hat{n})\cdot a_s'(\hat{n})\cdot\partial_x\phi\right) \tag{21}$$
$$+\phi(1-\phi^2) - \lambda(1-\phi^2)^2\left[U+\frac{z-z_0-\int V_P(t)dt}{l_T}\right]$$

$$\left(\frac{1+k}{2} - \frac{1-k}{2}\phi\right)\frac{\partial U}{\partial t} = \nabla\cdot\left[D_L\cdot q(\phi)\cdot\vec{\nabla}U - \vec{j}_{at}\right] + \frac{1}{2}\left[1+(1-k)U\right]\frac{\partial\phi}{\partial t} \tag{22}$$

where,

$$l_T = \frac{\Delta T_0}{G(t)} = \frac{|m|c_\infty(1-k)}{kG(t)}$$

In the equations, $D_L$ is the diffusion coefficient in the liquid. $l_T$ is the thermal length, where m is the slope of the liquidus line in the phase diagram. The interpolation function q(ϕ)=(1−ϕ)/2 determines the varied diffusion coefficient across the whole domain. Neglecting the effect of kinetic undercooling, the calculation parameters in the PF equations can be linked to the physical qualities by $W=d_0\lambda/a_1$ and $\tau_0=a_2\lambda W^2/D_L$, where W and $\tau_0$ are the interface width and relaxation time, which are the length scale and time scale, respectively. In the expressions, $a_1=5\sqrt{2}/8$, $a_2=47/75$, $\lambda$ is the coupling constant, $d_0=\Gamma/|m|(1-k)(c_\infty/k)$ is the chemical capillary length. $\Gamma=\gamma_{sl}T_f/(\rho_s L_f)$ is the Gibbs-Thomson coefficient, where $\gamma_{sl}$ is surface energy between the solid and liquid, $T_f$ is the melting point of pure solvent and $L_f$ is the latent heat, respectively.

## 2. Simulation procedure

The material Al-2.0wt.%Cu could be regarded as a dilute binary alloy, whose material parameters are shown in Table 1. [40,41]

When solving the PF equations, the most important calculation parameter is the interface width W. [42] The accuracy of simulation increases with the decrease of W, while the computational cost increases dramatically with the decrease of W. Through implementing the thin interface limitation in the PF model, the magnitude of W just needs to be one order of magnitude smaller than the characteristic length scale of the structures. [24,43] The characteristic length of alloy solidification is $L_C \sim \sqrt{d_0}*D_L/V_{tip}$,[3] hence W was set to be 0.15μm. During the simulation, the periodic boundary conditions were loaded for the phase field and supersaturation field along the Thermal Gradient Direction (TGD). The time step size was chosen below the threshold of numerical instability for the diffusion equations, i.e., $\Delta t<(\Delta x)^2/(4D_L)$. Finally, this study used fixed grid size $\Delta x=0.8W$ and time step size $\Delta t=0.013\tau_0$.

Table 1. The material parameters of Al-2.0wt.%Cu for the simulation [40,41]

| Symbol | Value | Unit |
| --- | --- | --- |
| Liquidus temperature, $T_L$ | 927.8 | K |
| Solidus temperature, $T_S$ | 896.8 | K |
| Diffusion coefficient in liquid phase, $D_L$ | $3.0\times10^{-9}$ | m$^2$/s |
| Equilibrium partition coefficient, k | 0.14 | / |
| Alloy composition, $c_\infty$ | 2.0 | wt.% |
| Liquidus slope, m | −2.6 | K/wt.% |
| Gibbs-Thomson coefficient, $\Gamma$ | $2.4\times10^{-7}$ | K·m |
| Anisotropic strength of surface energy, $\varepsilon_4$ | 0.01 | / |

Moreover, to consider the infinitesimal perturbation of thermal noise on the S/L interface, a fluctuating current $J_U$ is introduced to the diffusion equation. By using the Euler explicit time scheme:

$$U^{t+\Delta t} = U^t + \Delta t \left( \partial_t U - \vec{\nabla} \cdot \vec{J}_U \right) \tag{23}$$

During the numerical simulation, the discretized noise in the 2D system becomes: [37,44]

$$\vec{\nabla} \cdot \vec{J}_U \approx \left( J_{x,i+1,j}^n - J_{x,i,j}^n + J_{y,i,j+1}^n - J_{y,i,j}^n \right) / \Delta x \tag{24}$$

The components of $J_U$ are random variables obeying a Gaussian distribution, which has the maximum entropy relative to other probability distributions: [36]

$$\left\langle J_U^m(\vec{r},\vec{t}) J_U^n(\vec{r}\,',\vec{t}\,') \right\rangle = 2D_L q(\phi) F_U^0 \delta_{mn} \delta(\vec{r}-\vec{r}\,') \delta(t-t') \tag{25}$$

In equation (25), the constant noise magnitude $F_u^0$ means the value of $F_U$ for a reference planar interface at temperature $T_0$, defined as: [37,44]

$$F_U^0 = \frac{k v_0}{(1-k)^2 N_A c_\infty} \tag{26}$$

where $v_0$ is the molar volume of the solute atom, and $N_A$ is the Avogadro constant. Using the Clausius-Clapeyron relation:

$$\frac{|m|}{1-k} = \frac{k_B T_0^2}{\Delta h} \tag{27}$$

where $\Delta h$ is the latent heat per mole, and $k_B$ is the Boltzmann constant. Then, the constant noise amplitude becomes:

$$F_U^0 = \frac{k}{|m|c_\infty(1-k)} \frac{k_B T_0^2}{L} \tag{28}$$

Finally, the program codes of the WL model and PF model were written by C++. The explicit Finite Difference Method (FDM) was used when solving the PF governing equations, and the Message Passing Interface (MPI) parallelization was adopted for improving the computational efficiency.

## III. RESULTS AND DISCUSSION

During directional solidification, the temperature decreases at the front of the interface, solidification takes place to maintain the local thermodynamic equilibrium. The planar instability appears firstly during the evolution of solidification, represented by the transition from the planar to the cellular. The instability affects the subsequent stages significantly, which deserves systematic investigation.

## A. The dynamic evolution of the planar instability

In this section, the WL model and PF model are adopted to simulate the crystal growth, for which the thermal gradient G is 100K/mm and the pulling speed $V_P$ is 300μm/s. To represent the same parameters in the WL model and PF model, the PCO in the PF simulation is set parallel with the TGD, i.e., $\theta_0=0°$. The computational domain of the PF simulation is 2400×2400 grids, corresponding to 288.0μm×288.0μm in the real unit. It takes about 30 hours using 40 cores to finish one PF program.

The evolution of the S/L interface and solute field at the initial growth stage is shown in Fig. 2. At this stage, the interface keeps planar and advances slowly to the liquid, with the accumulation of the solute ahead of the interface. As time goes on, the planar instability appears, represented by the transition from the planar to the cellular, shown in Fig. 2(d)-(e). According to Fig. 2(d), the crossover time of the planar instability is about t=0.39s.

Then, the evolution of the characteristic parameters is extracted from the PF model and compared with the WL model, in Fig. 3(a), including the concentration ahead of the interface $c_0$ and the instantaneous velocity of the interface $V_I$. In Fig. 3(a), before the crossover time (t=0.39s), the curves from the WL model and PF model show good consistencies with each other, validating the accuracy of the PF model. After the crossover time, the curves extracted from the WL model and PF model differ from each other. Because the parameters from the WL model are based on the planar interface, the complex morphologies of the interface caused by the instability give rise to the differences between the WL model and PF model after the crossover time. That is, the differences do not mean the contradictions between these two models.

In Fig. 3(a), the solute concentration $c_0$ increases with time, and so does the instantaneous velocity $V_I$. In the PF simulation, on the one hand, at the crossover time, the cellular appearing increases the instantaneous velocity significantly, shown by the sharp increment of the $V_I$ curve of the PF model in Fig. 3(a). On the other hand, the solute at the S/L interface should satisfy the conservation law. After the crossover time, the solute still accumulates ahead of the interface, shown by the increment (limited) of the $c_0$ curve of the PF model after the crossover time in Fig. 3(a). After the cellular appearing, rather than diffusing only along the pulling direction of the planar interface, the solute could diffuse along multiple directions from the cellular tip to the liquid. As a result, the solute concentration starts to decrease, shown by the decrease of the $c_0$ curve of the PF model after the peak (about t=0.43s) in Fig. 3(a). The distributions of solute across the interface at different times, at the initial growth stage, are extracted from the PF simulations and shown in Fig. 3(b). On the one hand, due to the solute trapping, the turning points of the curves mean the positions of the interface. On the other hand, the peaks of the curves in Fig. 3(b) represent the solute concentrations ahead of the interface, which show the same tendency as compared with the $c_0$ curve of the PF model in Fig. 3(a).

In conclusion, at the initial stage, both the instantaneous velocity and concentration ahead of the interface increase with time. The sharp increment of the instantaneous velocity reflects the onset time of the planar instability, and the peak of the solute concentration reflects the completion of the cellular appearing.

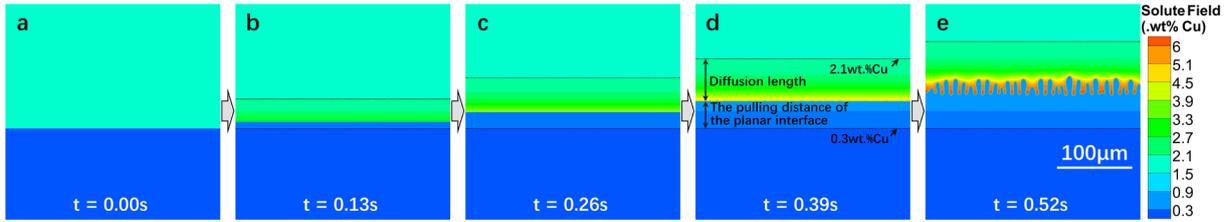

Fig. 2. The evolution of the interface and solute field at the initial growth stage. (from the PF simulations)

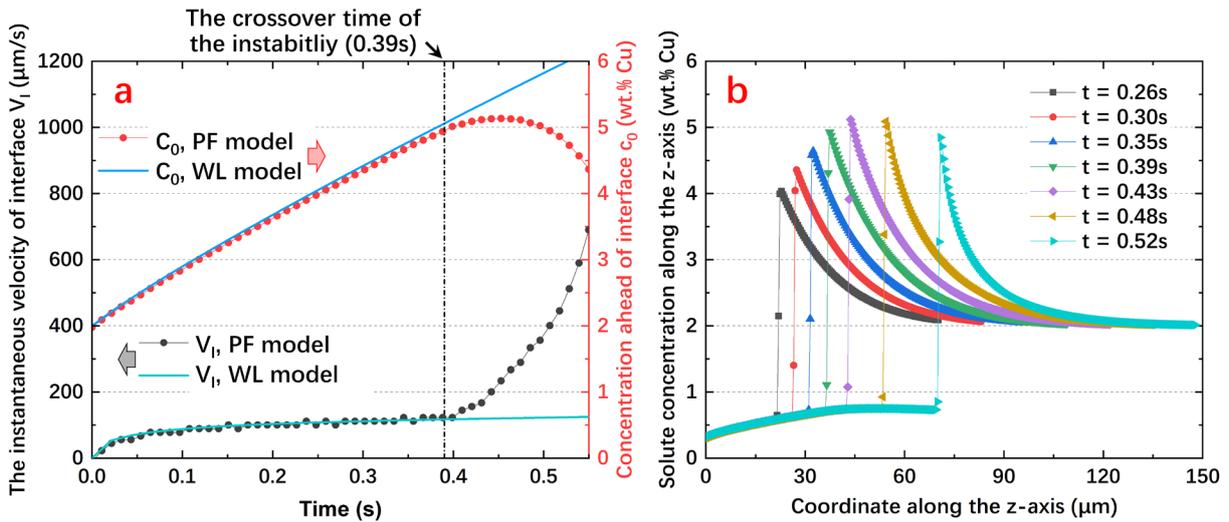

Fig. 3. The evolution of the characteristic parameters: (a) the evolution of the concentration ahead of the interface $c_0$ and the instantaneous velocity of the interface $V_I$ (from the WL model and PF model); (b) the distributions of solute across the interface at different times at the initial stage. (from the PF simulations)

## B. The criterion of the planar instability under varying conditions

### 1. The determination of the criterion

Due to the importance of the planar instability, the criterion of whom need to be determined. Hunt and Lu [45] consider the thermal gradient G and pulling speed $V_P$ as the determining parameters of microstructure selection. Based on G and $V_P$, they sketch a microstructure selection map. Brener et al. [46,47] construct a kinetic phase diagram, based on the effective anisotropy and undercooling, representing the regions of existence of different structures and the lines of transitions between the structures. However, neither Hunt's nor Brener's viewpoint considers the history-dependence of solidification. Their selection maps can hardly predict the evolution at the initial stage accurately. According to the WL analysis, the solidification evolution depends on the detailed way the conditions are achieved, including the planar instability and the subsequent stages. [11] In this section, the planar instability under the varying conditions is investigated.

Firstly, the evolution under steady-state and varying conditions is compared. Under the steady-state conditions, G is 100K/mm and $V_P$ is 300μm/s all the time, shown by the blue curve in Fig. 4(e). By contrast, under the varying conditions, G is constant while $V_P$ increases from 0 to a fixed value of 300μm/s. The increase times of $V_P$ are 0.5s, 1.0s, 2.0s and 3.0s, respectively, shown by the blue curves in Fig. 4(a)-(d). The characteristic parameters is shown in Fig. 4(a)-(e), including the concentration ahead of the interface $c_0$ and the instantaneous velocity of the interface $V_I$. Fig. 4 illustrates the $V_I$ curves and $c_0$ curves show similar tendencies with each other, including the sharp increment of the $V_I$ curves caused by the planar instability and the decrease of the $c_0$ curves caused by the cellular appearing. Meanwhile, the quantitative features of these curves differ from each other. Specifically, under the constant G and $V_P$, in Fig. 4(e), the acceleration rate of $V_I$ decreases with time, while the increase rate of $c_0$ is constant. The $V_I$ curve is logarithmic-function-like, while the $c_0$ curve is linear-function-like, consistent with literature [12]. By contrast, under the constant G and varying $V_P$, in Fig. 4(a)-(d), at the initial stage, the acceleration rate of $V_I$ is almost constant, while the increase rate of $c_0$ increases with time. The $V_I$ curve is linear-function-like, while the $c_0$ curve is exponential-function-like, consistent with literature [15,44]. On the one hand, the results demonstrate the distinctions between the steady-state conditions and varying conditions. Since it consumes time to adjust the parameters in practice, it is more suitable to use the varying parameters. On the other hand, the different evolutional characteristics between the $V_I$ curves and the $c_0$ curves indicate the driving force of the interface is not only determined by the solute diffusion, which will be investigated in detail in the future.

Back to the criterion of the planar instability. In the subsequent discussion, the varying solidification parameters are used, in Fig. 4(a)-(d), the increase times of $V_P$ are 0.5s, 1.0s, 2.0s and 3.0s, respectively. The corresponding evolution of the solute concentration $c_0$ and interface velocity $V_I$ is also shown in Fig. 4(a)-(d). As mentioned before, the sharp increment of the $V_I$ curve represents the onset time of the planar instability. With the acceleration rate of $V_P$ decreasing, the crossover time of the planar instability increases, shown in Fig. 4 from (a) to (d). The phenomenon demonstrates the conclusion the solidification evolution depends on the detailed way the conditions are achieved. Moreover, the planar instability may occur before the pulling speed $V_P$ becomes constant, shown in Fig. 4(b)-(d). That is, under the varying conditions, the crossover times of the instability correspond to different values of $V_P$, indicating that the thermal parameters G and $V_P$ are not the critical parameters of the interfacial instability.

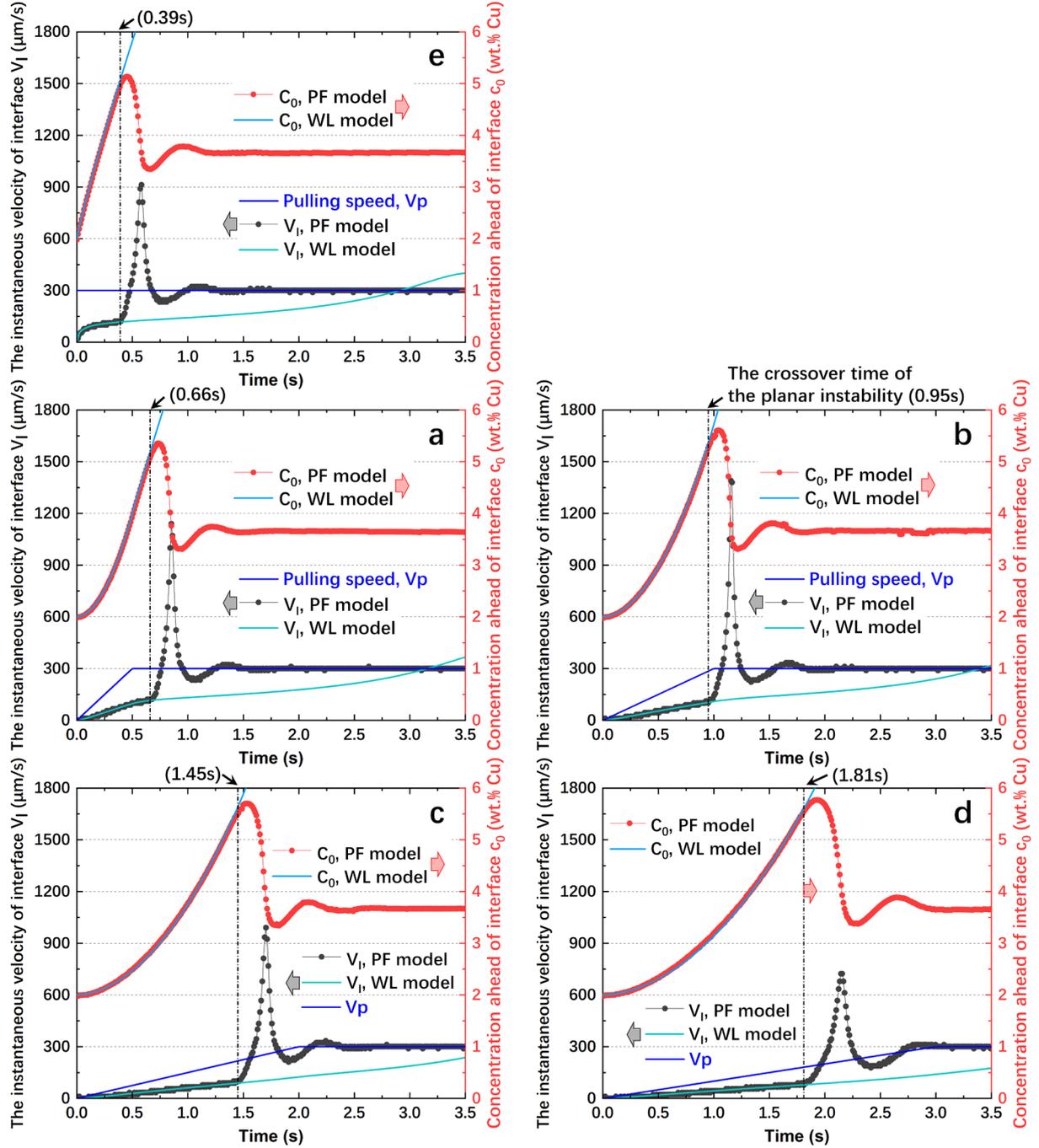

Fig. 4. The evolution under varying conditions: The constant G and varying $V_P$ (the increase times of $V_P$ are 0.0s, 1.0s and 2.0s, respectively) and the corresponding evolution of the concentration ahead of the interface $c_0$ and the instantaneous interfacial velocity $V_I$. (from the WL model and PF model)

Here, rather than the thermal parameters G and $V_P$, this paper considers the critical parameters of the instability are the excess free energy at the interface and the corresponding interfacial energy, as following:

The atoms at the S/L interface need to accommodate the slight structural changes on both solid and liquid sides, bringing the excess free energy. The integral of the excess free energy of the interface, multiplied by molar volume $V^m$, is the S/L interfacial energy (unit: J/m$^2$): [3]

$$\gamma_{sl} = \frac{1}{V^m} \int \Delta G^m(z) dz \quad (29)$$

Equation (29) illustrates the interfacial energy is determined by the excess Gibbs free energy (when the pressure is ignored, the Gibbs free energy equals to the Helmholtz free energy). The material is regarded as the binary alloy made up of solvent Al and solute Cu. Taking the liquid as the reference state (where $G_{Al}^m$ and $G_{Cu}^m$ are zero), the molar free energies of the liquid and solid are given by:

$$G_L^m(X_L, T) = RT\left[X_L \ln X_L + (1-X_L)\ln(1-X_L)\right] + \Omega_L X_L(1-X_L) \quad (30)$$

$$\begin{aligned} G_S^m(X_S, T) = &(1-X_S)\Delta S_f^{Al}(T-T_f^{Al}) + X_S \Delta S_f^{Cu}(T-T_f^{Cu}) \\ &+ RT\left[X_S \ln X_S + (1-X_S)\ln(1-X_S)\right] + \Omega_S X_S(1-X_S) \end{aligned} \quad (31)$$

where R is the gas constant, T is the temperature. $X_L$ is the molar composition of Cu at the liquid side, i.e., the concentration ahead of the interface. $X_S$ is the molar composition of Cu at the solid side. $\Delta S_f^{Al}$ and $\Delta S_f^{Cu}$ are the fusion entropies of pure Al and Cu, respectively. $T_f^{Al}$ and $T_f^{Cu}$ are the fusion points of pure Al and Cu, respectively. $\Omega_L$ and $\Omega_S$ are the regular solution model parameters.

When solidification takes place, the free energy reduces from $G_L$ to $G_S$. The difference between $G_L$ and $G_S$ is the driving force for solidification $\Delta G$. According to the local equilibrium approximation, $X_S$ and $X_L$ satisfy the relationship $X_S = k \cdot X_L$. Finally, the driving force $\Delta G$ is given by:

$$\begin{aligned} \Delta G &= G_L^m(X_L, T) - G_S^m(X_S, T) \\ &= \Omega_L X_L(1-X_L) - \Omega_S k X_L(1-kX_L) \\ &\quad - (1-kX_L)\Delta S_f^{Al}(T-T_f^{Al}) - kX_L \Delta S_f^{Cu}(T-T_f^{Cu}) \\ &\quad + RT\left[X_L \ln X_L + (1-X_L)\ln(1-X_L) - kX_L \ln(kX_L) - (1-kX_L)\ln(1-kX_L)\right] \end{aligned} \quad (32)$$

According to equation (32), we know $\partial \Delta G / \partial T < 0$ and $\partial \Delta G / \partial X_L < 0$ (when $X_L < X_{eutectic}$), illustrating $\Delta G$ has negative relations with both T and $X_L$. Moreover, the abstract of $\partial \Delta G / \partial X_L$ is much larger than that of $\partial \Delta G / \partial T$, meaning the effect of $X_L$ is much greater than T. As solidification goes on, at the S/L interface, T decreases and $\Delta G$ increases, meanwhile $X_L$ increases and $\Delta G$ decreases. Because the effect of $X_L$ is much greater than T. As time goes by, the excess free energy $\Delta G$ decreases, so does the corresponding interfacial energy $\gamma_{sl}$. When $\gamma_{sl}$ reduces to a critical level, with the influence of the interfacial anisotropy, the planar instability appears.

In conclusion, the change of the excess free energy at the interface, caused by the transport processes of heat and solute, and the corresponding interfacial energy are the critical parameters of the planar instability.

**2. The validation of the criterion**

**(1) Compared with the theoretical model**

To validate the conclusion in section III.B.1, the crossover times of the planar instability from the PF model are compared with those from the theoretical model. By solving equations (12)-(13), the increase rate of the perturbation $\sigma_\omega$ and the amplitude of the perturbation $A_\omega$ can be obtained. The varying solidification parameters are used. The thermal gradient G is constant 100K/mm, while the pulling speed $V_P$ increases from 0 to a fixed value of 300μm/s, for which the increase time is 2.0s.

Based on the theoretical model, the increase rates $\sigma_\omega$ and the corresponding perturbation amplitudes $A_\omega$ under the varying conditions are shown in Fig. 5. The times when $\sigma_\omega$ becomes positive, shown by the red curves in Fig. 5(a1)-(d1), representing the critical times of the marginal stability. It should be pointed out, the critical time $t_c$ here just reflects the time that the perturbations can be amplified. At this moment, the perturbations at the interface are still infinitesimal, which cannot be observed at the microscale. Hence, rather than the time when $\sigma_\omega$ turns positive, we define the crossover time of the instability based on a specific value of $A_\omega$. The time when the magnitude of $A_\omega$ reaches about 1μm is identified as the crossover time of the instability, shown by the purple curves in Fig. 5(a2)-(d2). In Fig. 5, the crossover times calculated from the theoretical model are 0.67s, 0.97s, 1.44s, and 1.83s, agreeing well with those from the PF simulations, 0.66s, 0.95s, 1.45s, and 1.81s, shown in Fig. 4(a)-(d).

It should be noted, in the theoretical model, based on the linear instability analysis, the perturbation is represented by the dispersion relation, in equation (12), defined as the increase rate of the perturbation as a function of wavenumber. By contrast, in the PF model, the perturbation is reflected by the noise term in the diffusion equation, shown in equations (23)-(28). During the construction of the PF model, the method of matched asymptotic expansions is adopted. The perturbation analyses have been carried out on each scale, including the inner scale (interface) and the outer scale (sharp-interface problem), then the two expansions are matched. [19] In this way, the dynamics of the interface in the PF model has been precisely controlled. The consistency between the two different methods validates the accuracy of the models.

**(2) Simulations of the crystals with different PCOs**

In addition, the solidification processes of the crystals with different PCOs are carried out. On the one hand, in the actual solidification procedure, the <100> directions of the crystals are not always parallel with the TGDs. On the other hand, due to the anisotropy, the crystals with different PCOs have different interface energies. In a 2D system, the anisotropic interfacial energy is given by:

$$\gamma_{sl} = \gamma_{sl}^0 \left[1 + \varepsilon_4 \cos 4(\theta + \theta_0)\right] \quad (33)$$

where $\gamma_{sl}^0$ is the value of the isotropic interfacial energy, determined by the excess free energy at the interface. $\varepsilon_4$ is a measure of the anisotropic strength, and $\theta_0$ is the intersection angle between the PCO and TGD. It should be noted, equation (33) only makes sense under the conditions the interfacial curvature exists. When the interfaces are planar, the PCO of the crystal does not affect the value of its interfacial energy. That is, the solidification evolution of the crystals with different PCOs could be used to test the conclusion about the influence of the free energy on the planar instability.

Due to the cubic structure of the Al-Cu alloy, the intersection angles between the PCO and TGD are set to be 0°, 15°, 30°, and 45°. The evolution of the characteristic parameters is shown in Fig. 6, including the concentration ahead of the interface $c_0$ and the instantaneous velocity of the interface $V_I$. Before the crossover time of the instability, the $c_0$ curves of the crystals with different PCOs overlap with each other completely, and so do the $V_I$ curves. The results indicate the interfacial anisotropy do not influence the transport processes at the planar growth stage, consistent with the literature [44]. As time goes on, the planar instability appears, represented by the sharp increase of the $V_I$ curves. Fig. 6 illustrates the crossover times of the instability are the same between the simulations, i.e., the PCO of the crystal has little effect on the crossover time of the instability. The phenomenon demonstrate the conclusion that the excess free energy $\Delta G$ and the corresponding interfacial energy $\gamma_{sl}^0$ are the critical parameters of the interfacial instability. Specifically, at the planar growth stage, the characteristic parameters of the crystals with different PCOs are the same, in Fig. 6, including the solute concentration $c_0$ and instantaneous velocity $V_I$. As a result, the magnitudes of $\Delta G$, dominated by the concentration and temperature, are the same between them. Because $\gamma_{sl}^0$ is determined by $\Delta G$, the changes of $\gamma_{sl}^0$ are also the same between the crystals with different PCOs. When the magnitude of $\gamma_{sl}^0$ decreases to the critical level, the planar instability occurs. Hence the crossover times are also the same between the crystals with different PCOs.

In conclusion, the comparisons between the PF model and theoretical model demonstrate the influence of the free energy on the planar instability. The simulations of the crystals with different PCOs also validate the conclusion the excess free energy and the corresponding interfacial energy are the critical parameters of the planar instability.

It needs to be noted, although the crossover times of the planar instability are the same in the simulations with different PCOs, the detailed evolutional characteristics of the planar-cellular-transition show differences, including the solute concentration $c_0$ and instantaneous velocity $V_I$, shown by the curves after the crossover time in Fig. 6. Differing from the planar growth stage, the interfacial curvature and its anisotropy influence the evolution significantly at this stage, which will be studied in the future.

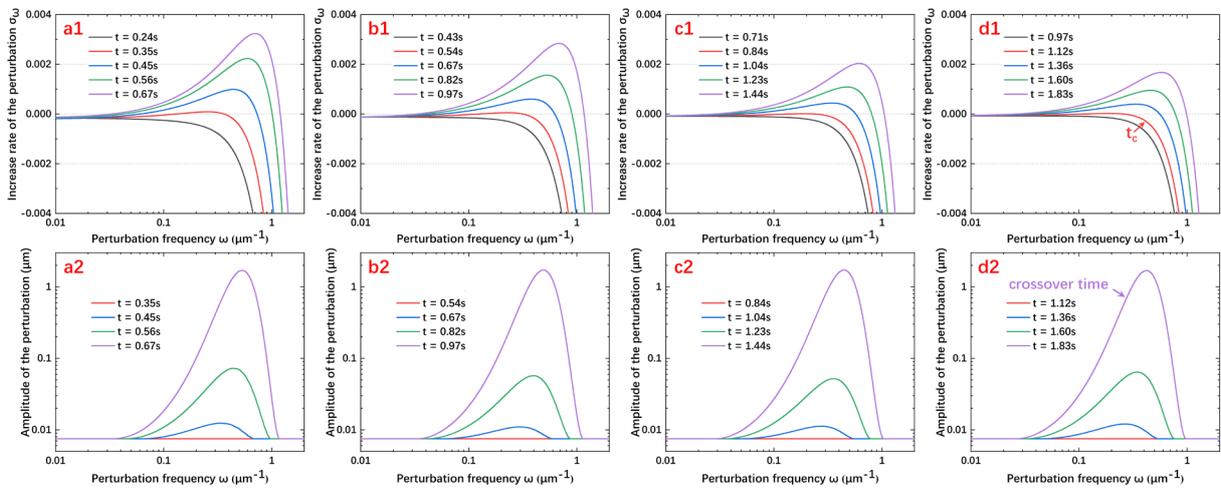

Fig. 5. Dynamic evolution of the planar instability predicted by the theoretical model: The evolution of the increase rate of the amplitude spectrum and the evolution of the amplitude spectrum. (The constant G and varying $V_P$, where the increase times of $V_P$ are 0.5s, 1.0s, 2.0s and 3.0s, respectively)

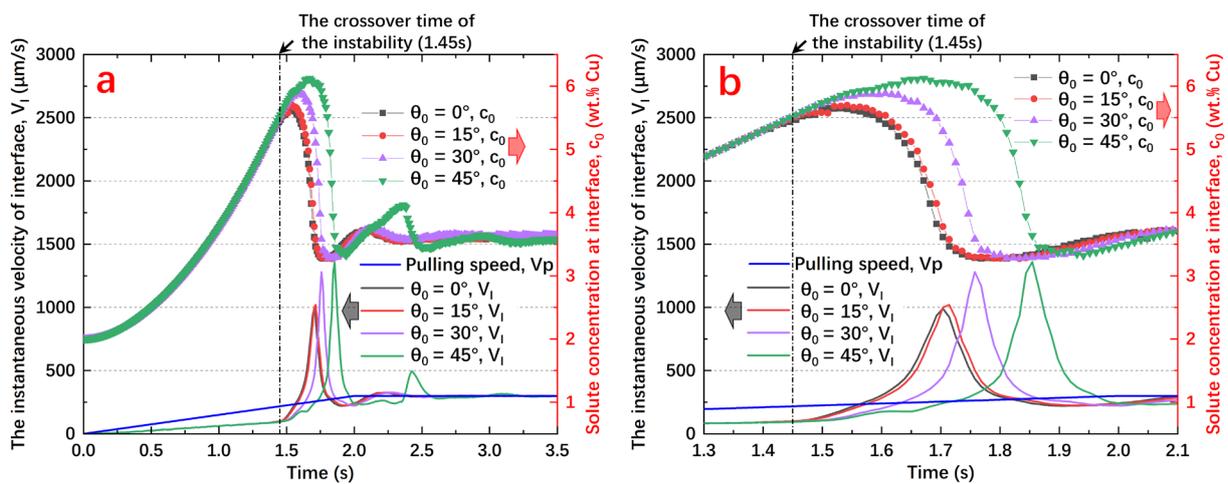

Fig. 6. The evolution of characteristic parameters with different PCOs: (a) the concentration ahead of the interface $c_0$ and the instantaneous interfacial velocity $V_I$; (b) the enlarged version of (a). (from the PF simulations)

## IV. CONCLUSIONS

In this paper, the planar instability in alloy solidification under varying conditions is studied. Firstly, the evolution of the planar instability is carried out by the WL model and PF model. Secondly, to represent the history-dependence of solidification, the varying parameters are used in the simulations. Then the criterion of the planar instability is discussed. Finally, to validate of the criterion, the comparisons between the PF model and theoretical model are performed. Moreover, solidification processes with different PCOs are carried out. The following conclusions can be drawn from this study:

(1) At the initial stage, both the instantaneous velocity and concentration ahead of the interface increase with time. The sharp increment of the instantaneous velocity reflects the onset time of the planar instability, and the peak of the solute concentration reflects the completion of the cellular appearing.

(2) The change of the excess free energy at the interface, caused by the transport processes of heat and solute, and the corresponding interfacial energy are the critical parameters of the planar instability.

(3) The comparisons between the PF model and theoretical model demonstrate the influence of the free energy on the planar instability. The simulations of the crystals with different PCOs validate the conclusion the excess free energy and corresponding interfacial energy are the critical parameters of the instability.

The idea of the interfacial energy influencing the planar instability can be applied for investigating other patterns induced by the interfacial instability, which will be studied in the future.

## CONFLICTS OF INTEREST

The author declared that there is no conflict of interest.

## DATA AVAILABILITY

The sharing of the data in this paper is available upon request.